\documentclass[a4paper,11pt]{article}
\usepackage{pos}

\title{The chiral condensate at large $N$}

\author*[a]{Claudio Bonanno}
\author[a,b]{Pietro Butti}
\author[a]{Margarita Garc\'ia Per\'ez}
\author[a,c]{Antonio Gonz\'alez-Arroyo}
\author[d,e]{Ken-Ichi Ishikawa}
\author[e]{Masanori Okawa}

\affiliation[a]{Instituto de F\'isica T\'eorica UAM-CSIC, Calle Nicol\'as Cabrera 13-15,\\Universidad Aut\'onoma de Madrid, Cantoblanco, E-28049 Madrid, Spain}
\affiliation[b]{Departamente de F\'isica T\'eorica, Facultad de Ciencias and \\ Centro de Astropart\'iculas y F\'isica de Altas Energ\'ias (CAPA),  
	\\ Universidad de Zaragoza, Calle Pedro Cerbuna 12, E-50009, Zaragoza, Spain}
\affiliation[c]{Departamento de F\'isica Te\'orica, Universidad Aut\'onoma de Madrid,\\M\'odulo 15, Cantoblanco, E-28049 Madrid, Spain}
\affiliation[d]{Core of Research for the Energetic Universe,\\Graduate School of Advanced Science and Engineering,\\Hiroshima University, Higashi-Hiroshima, Hiroshima 739-8526, Japan}
\affiliation[e]{Graduate School of Advanced Science and Engineering, Hiroshima University,\\Higashi-Hiroshima, Hiroshima 739-8526, Japan}

\emailAdd{claudio.bonanno@csic.es}
\emailAdd{pbutti@unizar.es}
\emailAdd{margarita.garcia@csic.es}
\emailAdd{antonio.gonzalez-arroyo@uam.es}
\emailAdd{ishikawa@theo.phys.sci.hiroshima-u.ac.jp}
\emailAdd{okawa@hiroshima-u.ac.jp}

\abstract{We present results for the large-$N$ limit of the chiral condensate computed from twisted reduced models. We followed a two-fold strategy, one constiting in extracting the condensate from the quark-mass dependence of the pion mass, the other consisting in extracting the condensate from the mode number of the Dirac operator.}

\FullConference{The 40th International Symposium on Lattice Field Theory (Lattice 2023)\\
July 31$^{\text{st}}$ - August 4$^{\text{th}}$, 2023\\
Fermi National Accelerator Laboratory\\}

\usepackage{braket}
\usepackage{placeins}
\usepackage[font=small, labelfont=bf, textfont={small,it}]{caption}
\newcommand{\beq}{\begin{eqnarray}}
\newcommand{\eeq}{\end{eqnarray}}
\newcommand{\beqnn}{\begin{eqnarray*}}
\newcommand{\eeqnn}{\end{eqnarray*}}

\newcommand{\Tr}{\ensuremath{\mathrm{Tr}}}
\newcommand{\SU}{\ensuremath{\mathrm{SU}}}

\newcommand{\MS}{\ensuremath{\overline{\mathrm{MS}}}}
\newcommand{\mutwoGeV}{\mu = 2~\ensuremath{\mathrm{GeV}}}
\renewcommand{\P}{\ensuremath{\mathrm{P}}}

\newcommand{\A}{\ensuremath{\mathrm{A}}}
\newcommand{\R}{\ensuremath{\mathrm{R}}}
\newcommand{\W}{\ensuremath{\mathrm{W}}}
\newcommand{\PCAC}{\ensuremath{\mathrm{PCAC}}}
\newcommand{\TEK}{\ensuremath{\mathrm{TEK}}}

\begin{document}
\maketitle

\section{Introduction}

Lattice strategies to study the large-$N$ limit of gauge theories can be mainly divided into two broad classes:
\begin{itemize}
\item Standard approaches, namely, numerical results obtained from theories discretized on regular extended lattices for $N \lesssim 10$ are extrapolated towards $1/N \to 0$~\cite{Lucini:2012gg,Bali:2013kia,Bonati:2016tvi,Ce:2016awn,DeGrand:2016pur,Hernandez:2019qed,Bennett:2020hqd,Bonanno:2020hht,Athenodorou:2021qvs,Bonanno:2022yjr,DeGrand:2023hzz};
\item Reduced models~\cite{PhysRevLett.48.1063,BHANOT198247,GONZALEZARROYO1983174,PhysRevD.27.2397}, namely, calculations are performed for $N = O(100)$ or larger (i.e., practically already in the large-$N$ limit), with the lattice volume either very small or completely reduced to just a single point~\cite{Gonzalez-Arroyo:2010omx,Gonzalez-Arroyo:2012euf,Lohmayer:2013spa,Gonzalez-Arroyo:2014dua,GarciaPerez:2014azn,GarciaPerez:2015rda,Perez:2015ssa,Gonzalez-Arroyo:2015bya,Perez:2017jyq,GarciaPerez:2020gnf,Perez:2020vbn,Butti:2022sgy,Butti:PoS}.
\end{itemize}

These two different methods can be regarded as complementary. On one hand, the excellent agreement found for the large-$N$ limit of several observables obtained from these two different methods is highly non-trivial. On the other hand, since the finite-$N$ corrections of the reduced models are different from those of the standard approach, results from both methods can be combined to extract the actual $N$-dependence of a given observable.

This manuscript deals with the computation of the large-$N$ limit of an observable which plays an intriguing role both from the theoretical and the phenomenological point of view: the chiral condensate. The computation of this observable has been performed for $\SU(3)$ with a variety of lattice discretizations and fermion contents~\cite{Engel:2014eea,Boyle:2015exm,Wang:2016lsv,Alexandrou:2017bzk,Aoki:2017paw,ExtendedTwistedMass:2021gbo,Liang:2021pql,Bonanno:2023xkg}, but so far only few large-$N$ determinations have been given in the literature, either involving just one lattice spacing~\cite{Narayanan:2004cp,Hernandez:2019qed}, or presenting a preliminary study of the large-$N$ limit~\cite{DeGrand:2023hzz}. This proceeding reports on the main results of~\cite{Bonanno:2023ypf}, where the large-$N$ limit of the chiral condensate is computed using the Twisted Eguchi--Kawai (TEK) model~\cite{GONZALEZARROYO1983174,PhysRevD.27.2397}. In particular, we obtain the chiral condensate at large $N$ from the low-lying spectrum of the Dirac operator, and we perform controlled continuum and chiral extrapolations to provide a solid determination of this quantity in the large-$N$ limit. Perfectly compatible results are obtained from the quark mass dependence of the pion mass, as we will show in the following.

This paper is organized as follows: in Sec.~\ref{sec:setup} we briefly summarize our numerical setup; in Sec.~\ref{sec:res} we present and discuss the main results of~\cite{Bonanno:2023ypf} about the large-$N$ limit of the chiral condensate; finally in Sec.~\ref{sec:conclu} we draw our conclusions.

\section{Numerical setup}\label{sec:setup}

Below we briefly summarize our TEK lattice discretization, as well as the Giusti--L\"uscher method to extract the chiral condensate from the low-lying spectrum of the lattice Dirac--Wilson operator.

\subsection{Lattice discretization}

The pure-gauge TEK model is a matrix model where the dynamical degrees of freedom are $d=4$ $\SU(N)$ matrices. It can be thought of as the reduction on a single site lattice of an ordinary Wilson lattice Yang--Mills theory defined on a discretized torus with twisted periodic boundary conditions. The Wilson pure-gauge TEK action reads:
\beq
S_{\W}^{(\TEK)}[U] = -N b \sum_{\nu \ne \mu} z_{\nu\mu} \Tr\left\{U_\mu U_\nu U_\mu^\dagger U_\nu^\dagger\right\},
\eeq
where $1/b$ is the bare 't Hooft coupling, $U_\mu$ are the $\SU(N)$ $d=4$ link matrices, $N=L^2$, and $z_{\nu\mu} = z_{\mu\nu}^* = \exp\left\{i \frac{2\pi k(N)}{\sqrt{N}} \right\}$ ($\nu>\mu$) is the twist factor, with $k(N)$ an integer number co-prime with $\sqrt{N}\in\mathbb{Z}$. There is now plenty of theoretical and numerical evidence that this model reproduces the infinite-volume large-$N$ behavior of ordinary Yang--Mills gauge theories~\cite{Gonzalez-Arroyo:2012euf,Gonzalez-Arroyo:2015bya,Gonzalez-Arroyo:2014dua,GarciaPerez:2014azn,Perez:2017jyq,Perez:2020vbn}. Concerning Monte Carlo methods, gauge configurations were generated using the over-relaxation algorithm described in~\cite{Perez:2015ssa}.

In our work we do not consider any dynamical fermion, as in the large-$N$ limit the contributions of fundamental flavors is exactly zero. We will however consider one flavor of Dirac--Wilson fermions in the valence sector. In this case the lattice Dirac operator reads~\cite{Gonzalez-Arroyo:2015bya}:
\beq
D_\W^{(\TEK)}[U] = \frac{1}{2\kappa} - \frac{1}{2} \sum_{\mu}\left[(\mathbb{I}+\gamma_\mu) \otimes \mathcal{W}_\mu[U] + (\mathbb{I}-\gamma_\mu)\otimes \mathcal{W}_\mu^\dagger \right],
\eeq
with $\kappa$ the hopping parameter and $\mathcal{W}_\mu[U]\equiv U_\mu \otimes \Gamma_\mu^*$, where $\Gamma_\mu$ represent the twist eaters, satisfying $\Gamma_\mu \Gamma_\nu = z_{\nu\mu}^* \Gamma_\nu \Gamma_\mu$. 

\subsection{The Giusti--L\"uscher method}

The Banks--Casher relation equates the chiral condensate to the spectral density $\rho$ of the eigenmodes of the Dirac operator in the origin:
\beq
\frac{\Sigma}{\pi} = \lim_{\lambda\to 0}\lim_{m\to0}\lim_{V\to\infty} \rho(\lambda,m)
\eeq

Another physical quantity that is equivalent to $\rho$, but that is more convenient to be computed on the lattice, is the mode number of the massive Dirac operator:
\beq
\braket{\nu(M)} &\equiv& \braket{ \# \,\, \vert i\lambda+m \vert \le M} \\ &=& V \int_{-\Lambda}^{\Lambda} \rho(\lambda,m) d\lambda, \qquad \Lambda^2 \equiv M^2 - m^2.
\eeq

Being the mode number and the spectral density connected by an integral relation, it is clear that the Banks--Casher implies a linear behavior of $\braket{\nu(M)}$ as a function of $\Lambda$:
\beq
\braket{\nu(M)} = \frac{2}{\pi} V\Sigma \Lambda + o(\Lambda),
\eeq
where higher-order terms in $\Lambda$ are sub-leading in $1/N$~\cite{Smilga:1993abc,Osborn:1998qb}.

The Giusti--L\"uscher method~\cite{Giusti:2008vb} consists in obtaining the chiral condensate $\Sigma$ from a numerical lattice computation of the slope of the mode number of the Dirac operator as:
\beq
\label{eq:sigma_eff}
\Sigma^{(\mathrm{eff})}(m) &=& \frac{\pi}{2V} \sqrt{1-\left(\frac{m}{\overline{M}}\right)^2} \left[\frac{\partial \braket{\nu(M)}}{\partial M}\right]\Bigg\vert_{M=\overline{M}} \longleftarrow \text{slope of $\braket{\nu(M)}$ in $M=\overline{M}$},\\\nonumber
\\
\Sigma &=& \lim_{m\to 0} \Sigma^{(\mathrm{eff})}(m),
\eeq
where $\overline{M}$ is the point in which the slope is computed.

As a final comment, let us here stress that, within TEK models, the obtained results should be thought of as if they were obtained on a lattice with effective size $\ell = a L = a \sqrt{N}$. Therefore, the volume appearing in~\eqref{eq:sigma_eff} is given by $V = a^4 (\sqrt{N})^4 = a^4 N^2$.

\section{Results}\label{sec:res}

In this manuscript, we will mainly refer to results obtained for $N=289$, for which the $k$ parameter appearing in the twist factor defined in the previous section was chosen to be $k=5$. Since we expect the following large-$N$ scaling for the chiral condensate:
\beq
\Sigma(N) = N \left[\bar{\Sigma} + O\left(\frac{1}{N^2}\right)\right],
\eeq
in the following we will always report results for $\Sigma/N$, which we expect to approach a finite large-$N$ limit. Finally, all reported results for the renormalized chiral condensate are always expressed in the $\MS$ scheme at the conventional renormalization scale $\mutwoGeV$.

\subsection{Best fit of the mode number}

\begin{figure}[!htb]
\centering
\includegraphics[scale=0.44]{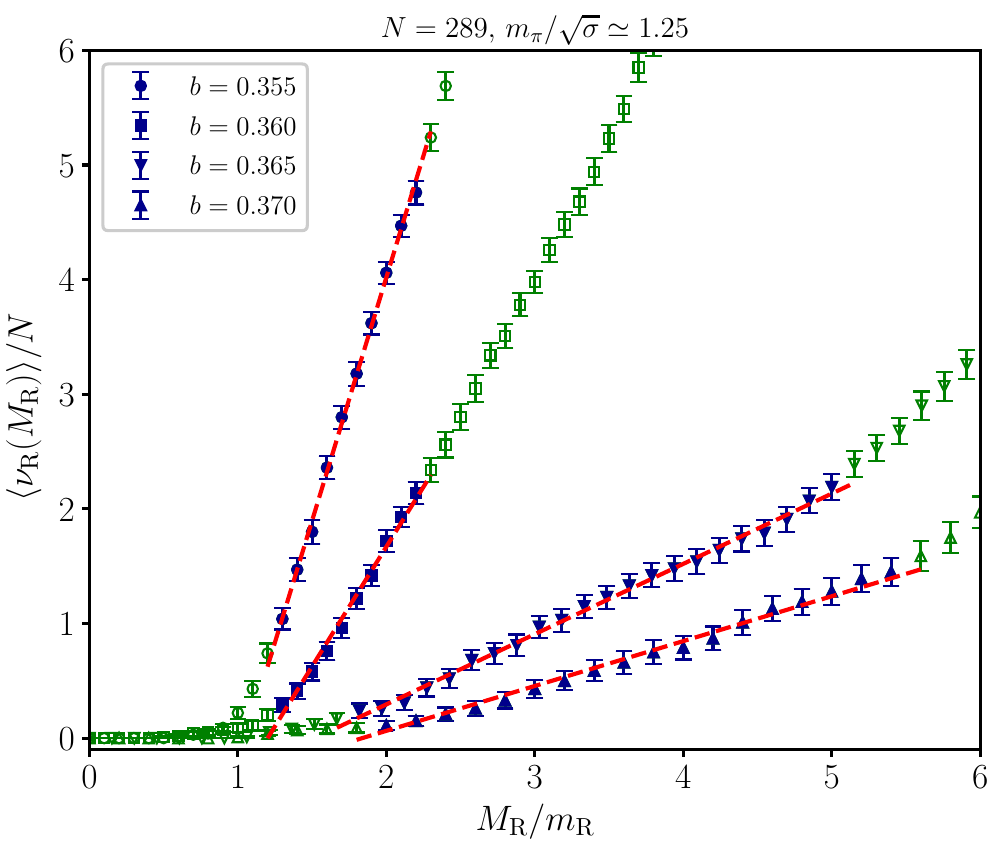}
\includegraphics[scale=0.45]{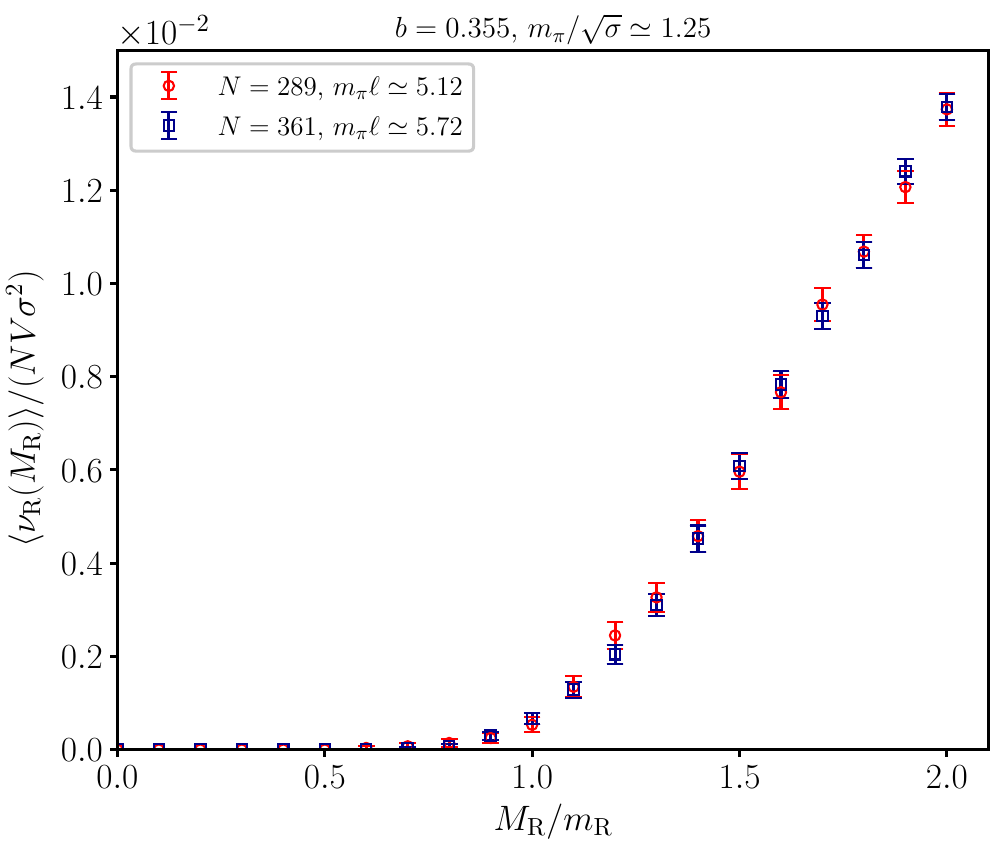}
\caption{Linear best fit of the mode number $\braket{\nu_\R}/N$ as a function of $M_\R/m_\R$ for 4 different values of the 't Hooft coupling $1/b$, and for 4 different values of $\kappa$ tuned to correspond approximately to the same value of the pion mass $m_\pi$ (top plot). Filled points correspond to the fitted ones. All our results refer to $N=289$, but in one case (bottom plot) we also checked that results obtained for $N=361$ where perfectly agreeing (in the latter case, $k=7$ was used for the twist factor).}
\label{fig:mode_number}
\end{figure}

Let us summarize our practical numerical implementation of the Giusti--L\"uscher method, as well as the procedure we followed to compute the chiral condensate from our Dirac spectra:
\begin{itemize}
\item Renormalization properties: $\braket{\nu}=\braket{\nu_\R}$, $\quad M_\R= M / Z_\P$, $\quad \lambda_\R = \lambda/Z_\P$;
\item Scale setting: determinations of the string tension $a \sqrt{\sigma}$ obtained from the TEK model in Refs.~\cite{Gonzalez-Arroyo:2012euf,Perez:2020vbn};
\item We solved numerically the eigenproblem $(\gamma_5 D^{(\TEK)}_{\mathrm{W}}[U]) u_\lambda = \lambda u_\lambda$ for 100 well-decorrelated gauge configurations using the \texttt{ARPACK} library, and computed the first 300 lowest-lying eigenvalues and eigenvectors;
\item From the knowledge of $Z_\A$ and $m_\PCAC$ (computed at large-$N$ from the TEK model in Ref.~\cite{Perez:2020vbn}), we obtained the renormalized eigenvalues as: $\lambda_\R/m_\R = \lambda/(Z_\A m_\PCAC)$;
\item We counted the renormalized modes $\lambda_\R/m_\R$ below the threshold $M_\R/m_\R$ to obtain $\braket{\nu(M_\R)} = \braket{\nu_\R(M_\R)}$;
\item From a linear best fit of $\braket{\nu_\R(M_\R)}/N$ vs $M_\R/m_\R$ we obtain the slope $\left[\frac{\partial \braket{\nu(M)}}{\partial M}\right]\Big\vert_{M=\overline{M}}$, where $\overline{M}$ is the middle point of the fit range (cf.~Fig.~\ref{fig:mode_number}, top panel);
\item From the slope of the mode number we extract the RG-invariant quantity $\Sigma^{(\mathrm{eff})}_\R m_\R/ (\sigma^2 N)$ from Eq.~\eqref{eq:sigma_eff};
\item Using $Z_\A m_\PCAC = Z_\P m_\R$ and the conventional value $\sqrt{\sigma} = 440$ MeV, we get rid of the quark mass and finally obtain the bare effective chiral condensate $\Sigma^{(\mathrm{eff})}_\R/( \sigma^{3/2} N Z_\P)$ in MeV$^3$;
\item All our results where obtained for $N=289$, but we explicitly checked that results obtained for $N=361$ gave perfectly agreeing results (cf.~Fig.~\ref{fig:mode_number}, bottom panel).
\end{itemize}

\FloatBarrier

\subsection{Chiral limit at fixed lattice spacing}

\begin{figure}[!htb]
\centering
\includegraphics[scale=0.42]{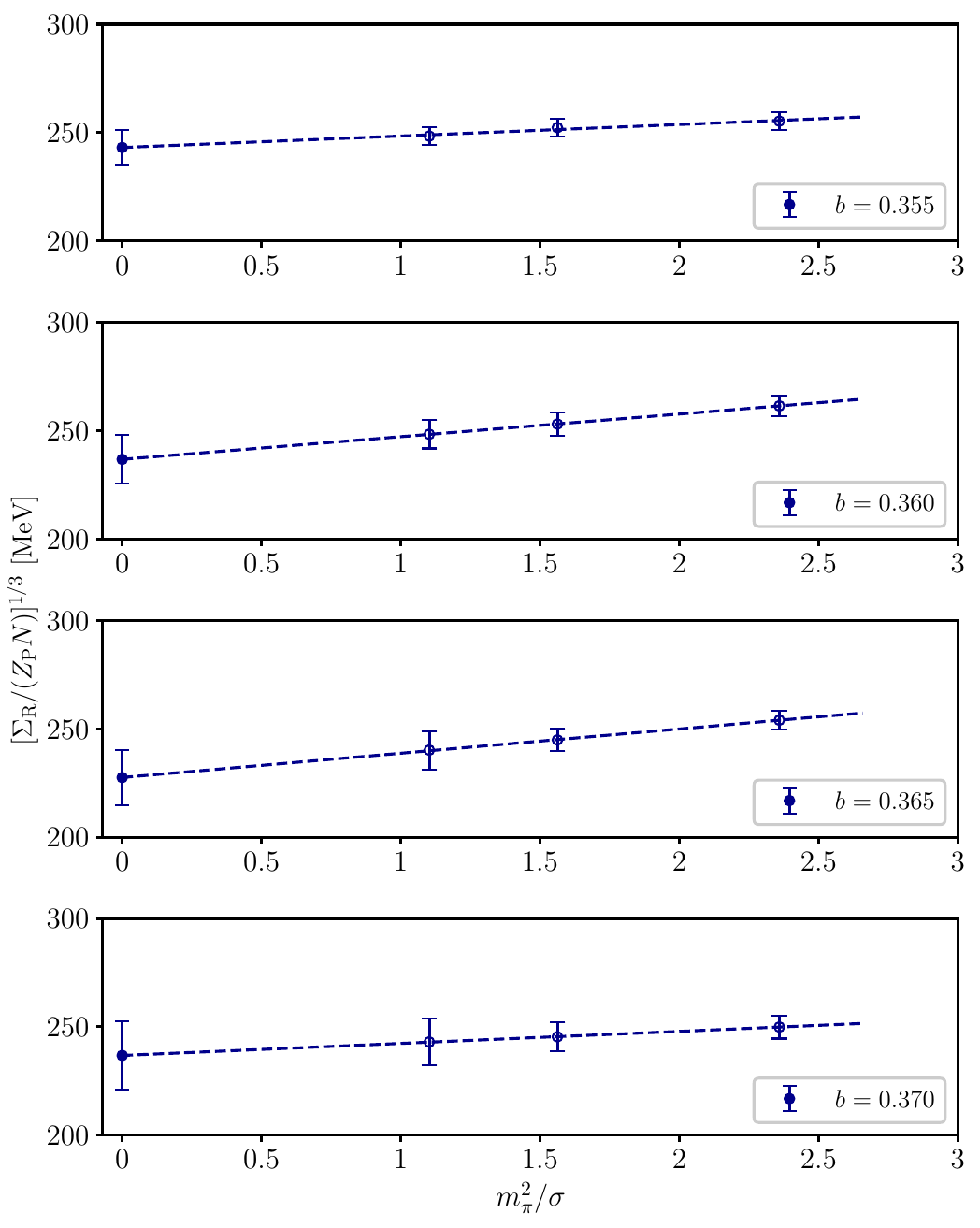}
\caption{Extrapolation towards the chiral limit of the bare effective chiral condensate $[\Sigma^{(\mathrm{eff})}_\R/(N Z_\P)]^{1/3}$ expressed in MeV physical units for all explored values of the 't Hooft coupling.}
\label{fig:chiral limit}
\end{figure}

After obtaining the bare effective chiral condensate from the Giusti--L\"uscher method, we need to extrapolate our determinations towards the chiral limit, in order to get rid of the finite quark mass we have used to determine our Dirac spectra. Our extrapolations towards the chiral limit, shown in Fig.~\ref{fig:chiral limit}, are done according to the Chiral Perturbation Theory (ChPT) prediction:
\beq\label{eq:ansatz_chiral_limit}
\Sigma^{(\mathrm{eff})}(m) &=& \Sigma + k\, m + o(m)\\
\nonumber\\[-1em]
&=& \Sigma + k^\prime \, m_\pi^2 + o(m_\pi^2),
\eeq
with $\Sigma^{(\mathrm{eff})} = \Sigma^{(\mathrm{eff})}_\R/ Z_\P$. Our chiral extrapolations will be performed, in all cases, at fixed $b$, i.e., at fixed lattice spacing. Thus, the renormalization constant $Z_\P$ will be, at fixed $b$, the same for all explored values of the pion mass $m_\pi$.

\FloatBarrier

\subsection{Continuum limit}

In order to extrapolate our results towards the chiral limit, we renormalize our large-$N$ determinations of $\Sigma_\R$ using the large-$N$ non-perturbative determinations of $Z_\P$ reported in~\cite{Castagnini:2015ejr}, where we refer the reader for more details about the numerical techniques to compute this renormalization constant from lattice simulations.

\begin{figure}[!htb]
\centering
\vspace*{-0.5cm}
\includegraphics[scale=0.43]{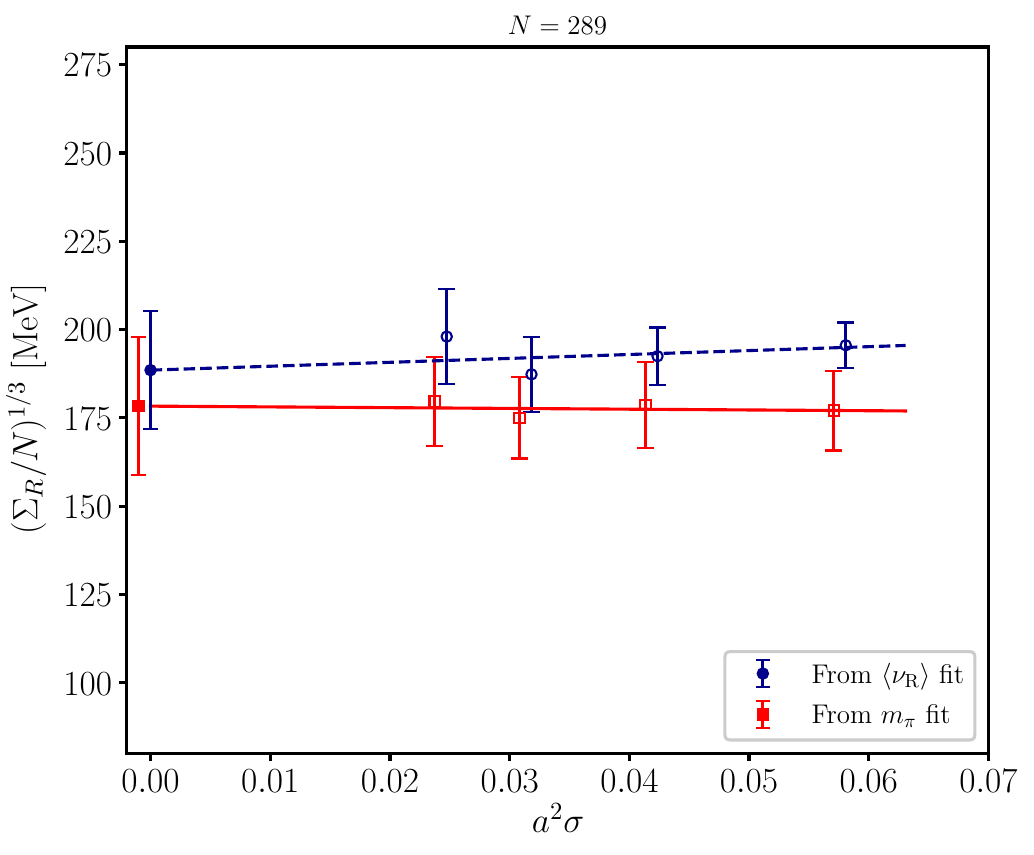}
\includegraphics[scale=0.43]{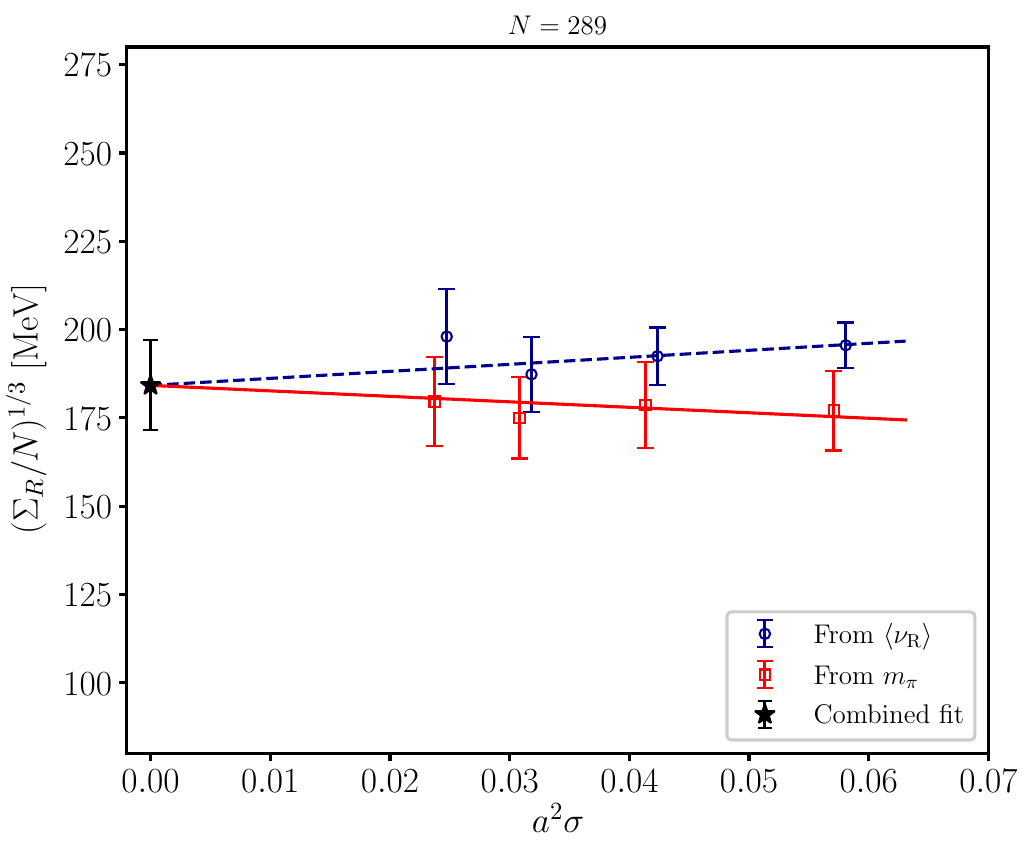}
\caption{Extrapolation towards the chiral limit of the spectral determinations of the chiral condensate obtained from the Giusti--L\"uscher method, compared with the determinations obtained from the GMOR relation~\eqref{eq:GMOR} and using the determinations of $m_\pi$ vs $m$ and of $F_\pi$ of Ref.~\cite{Perez:2020vbn}. In the top panel we show individual fits of the two data sets, in the bottom panel we show instead the combined fit of the two data sets.}
\label{fig:continuum limit}
\end{figure}

The continuum extrapolation of our spectral determinations of the chiral condensate, shown in Fig.~\ref{fig:continuum limit} (top panel), are done assuming leading $O(a^2)$ corrections as usual:
\beq
\Sigma_\R(a) = \Sigma_\R + c\, a^2 + o(a^2).
\eeq
In the top panel of Fig.~\ref{fig:continuum limit} we also show the renormalized determinations of the chiral condensate obtained from the Gell-Man--Oakes--Renner (GMOR) relation,
\beq\label{eq:GMOR}
m_\pi^2 = 2 \frac{\Sigma}{F_\pi^2}m = 2 \frac{\Sigma_\R}{F_\pi^2}m_\R,
\eeq
according to the large-$N$ TEK determinations of the pion mass and of the pion decay constant of Ref.~\cite{Perez:2020vbn}. We find perfectly agreeing results once the continuum limit is taken. Thus, we perform a combined continuum limit (see bottom panel of Fig.~\ref{fig:continuum limit}), giving the joint estimate:
\beq
(\Sigma_\R/N)^{1/3} = 184(13)~\mathrm{MeV}.
\eeq

\FloatBarrier

\section{Conclusions}\label{sec:conclu}

This manuscript reports on the main results of Ref.~\cite{Bonanno:2023ypf}, which presents a solid computation of the large-$N$ chiral condensate from TEK models using the Giusti--L\"uscher spectral method for $N=289$, using 4 lattice spacings and 3 pion masses each to provide controlled chiral and continuum extrapolations. The obtained results are in perfect agreement with those obtained from the quark mass dependence of the pion mass and the GMOR relation, giving a joint estimate of:
\beq
\lim_{N\to\infty} \frac{\Sigma_\R(N)}{N} = [184(13)\text{ MeV}]^3, \qquad \text{ ($\MS$, $\,\, \mutwoGeV$, $\,\, \sqrt{\sigma}=440$ MeV)}
\eeq

Our final result is in remarkable agreement with the FLAG21~\cite{FlavourLatticeAveragingGroupFLAG:2021npn} world-average for 2-flavor QCD $\Sigma_\R(N=3)/3 = [184(7)\text{ MeV}]^3$ when using $\sqrt{\sigma}$ to set the scale. Our calculation thus points out that $1/N^2$ corrections are small and $N=3$ is already very close to $N=\infty$. Such conclusion fits very well with other large-$N$ calculations pointing towards the same scenario~\cite{Lucini:2012gg,Bali:2013kia,Bonati:2016tvi,Ce:2016awn,Hernandez:2019qed,Bennett:2020hqd,Bonanno:2020hht,Athenodorou:2021qvs,Bonanno:2022yjr}.

In the next future, we plan to extend our calculation of the chiral condensate to the case of adjoint Majorana fermions, which is of great theoretical interest.

\section*{Acknowledgements}

This work is partially supported by the Spanish Research Agency (Agencia Estatal de Investigaci\'on) through the grant IFT Centro de Excelencia Severo Ochoa CEX2020-001007-S, funded by MCIN/AEI/10.13039 /501100011033, and by grant PID2021-127526NB-I00, funded by MCIN/AEI/10.13039/ 501100011033 and by “ERDF A way of making Europe”. We also acknowledge support from the project H2020-MSCAITN-2018-813942 (EuroPLEx) and the EU Horizon 2020 research and innovation programme, STRONG-2020 project, under grant agreement No 824093. P.~B.~is supported by Grant PGC2022-126078NB-C21 funded by MCIN/AEI/ 10.13039/501100011033 and “ERDF A way of making Europe”. P.~B.~also acknowledges support by Grant DGA-FSE grant 2020-E21-17R Aragon Government and the European Union - NextGenerationEU Recovery and Resilience Program on ‘Astrofísica y Física de Altas Energías’ CEFCA-CAPA-ITAINNOVA. K.-I.~I.~is supported in part by MEXT as "Feasibility studies for the next-generation computing infrastructure". M.~O.~is supported by JSPS
KAKENHI Grant Number 21K03576. Numerical calculations have been performed on the \texttt{Finisterrae~III} cluster at CESGA (Centro de Supercomputaci\'on de Galicia). We have also used computational resources of Oakbridge-CX at the University of Tokyo through the HPCI System Research Project (Project ID: hp230021 and hp220011).

\providecommand{\href}[2]{#2}\begingroup\raggedright\endgroup

\end{document}